\documentclass{article}

\usepackage{arxiv}

\usepackage[utf8]{inputenc} 
\usepackage[T1]{fontenc}    
\usepackage{hyperref}       
\usepackage{url}            
\usepackage{booktabs}       
\usepackage{amsfonts}       
\usepackage{nicefrac}       
\usepackage{microtype}      
\usepackage{graphicx}
\usepackage{lipsum}		

\def\cov{\mathrm{Cov} }
\def\var{\mathrm{Var} }
\def\diag{\mathrm{diag} }

\title{On Cokriging, Neural Networks, and Spatial Blind Source Separation for Multivariate Spatial Prediction}

\date{} 					

\author{Christoph~Muehlmann \\
	Vienna University of Technology \\
	\texttt{christoph.muehlmann@tuwien.ac.at} \\
	\And
	Klaus~Nordhausen \\
	Vienna University of Technology \\
	\texttt{klaus.nordhausen@tuwien.ac.at} \\
	\And
	Mengxi~Yi \\
	Vienna University of Technology and \\
	University of International Business and Economics, Beijing \\
	\texttt{myi@uibe.edu.cn} \\
}

\renewcommand{\headeright}{}
\renewcommand{\undertitle}{}

\begin{document}
\maketitle

\begin{abstract}
Multivariate measurements taken at irregularly sampled locations are a common form of data, for example in geochemical analysis of soil. In practical considerations predictions of these measurements at unobserved locations are of great interest. For standard multivariate spatial prediction methods it is mandatory to not only model spatial dependencies but also cross-dependencies which makes it a demanding task. Recently, a blind source separation approach for spatial data was suggested. When using this spatial blind source separation method prior the actual spatial prediction, modelling of spatial cross-dependencies is avoided, which in turn simplifies the spatial prediction task significantly. In this paper we investigate the use of spatial blind source separation as a pre-processing tool for spatial prediction and compare it with predictions from Cokriging and neural networks in an extensive simulation study as well as a geochemical dataset.
\end{abstract}


\section{Introduction}\label{sec:intro}

It is increasingly common that multivariate phenomena are connected to different spatial locations. Such data can be represented using a random vector {\small $X(s) = \left( X_1(s),\dots, X_p(s)\right) ^ \top$} indexed by continuous locations $s$ lying in a d-dimensional domain $\mathcal{S}^d \subseteq \mathbb{R} ^ d $. Such a collection $X(s)$ is also denoted as a multivariate random field. Usually, observations are only available at different, irregularly spaced, discrete sites $s_1,\dots,s_n$. A question of interest is what will be the values of $X$ at unobserved locations? Which means one is interested in multivariate spatial prediction. Key characteristics of multivariate spatial data are that there are dependencies between the different variables and also that observations taken in close proximity tend to be more similar than those taken far apart. Hence, it is natural that both sources of dependencies should be considered when predicting the values for new locations. As an example, \cite{MeyerEtAl2019} point out that the spatial nature of the data should not be ignored when using machine learning tools. This is also emphasized in \cite{Jiang2019}, a survey about spatial prediction in a univariate regression context.

In geostatistics, the spatial dependencies are usually considered by modelling the spatial covariance matrix $C(s_1,s_2)$ $=$ $\cov (X(s_1), X(s_2))$. But, as pointed out in a recent review by \cite{genton2015} it is quite challenging to define models and classes that are valid and flexible spatial covariance matrices. Also, the estimation of these matrices is quite demanding, thus, making multivariate spatial prediction a difficult task. Based on $C(s_1,s_2)$ the most popular forms of spatial prediction is Kriging for the univariate and Cokriging for the multivariate case. As an alternative to Kriging, neural networks have also been considered. However, the literature mainly focuses on prediction of univariate spatial data which is for example reviewed in \cite{LiHeap2013}.

Recently, \cite{sbss_comp,10.1093/biomet/asz079} suggest a blind source separation (BSS) model for spatial data which decomposes the $p$-variate random field $X(s)$ into $p$ independent univariate fields $Z_1(s),\dots,Z_p(s)$. This method can be used to simplify the task of multivariate prediction to independent univariate ones which in turn is much simpler than developing a multivariate model. The goal of this paper is to recap spatial blind source separation based on \cite{sbss_comp,10.1093/biomet/asz079} and describe the benefits of its use prior spatial prediction which has not yet been considered (Section~\ref{sec:sbss}) and compare this procedure to the commonly used Cokriging (Section~\ref{sec:lmc}) and neural networks (Section~\ref{sec:nn}) in an extensive simulation study (Section~\ref{sec:simu}) as well as in a real data example (Section~\ref{sec:real_data}).

For the remainder of the paper we assume that $X(s)$ is an isotropic and stationary random field which means that (i) $E(X(s))=\mu$, (ii) $\cov(X(s))=C_0$ and (iii) $C(s_1,s_2) = C(\| h \|)=C_h$ for all $s,s_1,s_2 \in \mathcal{S}$, where $h = s_1-s_2$. Thus, these assumptions imply that the first two moments are constant over all locations $s$ and the spatial covariance is only a function of the relative distances between locations.

\section{Linear model of coregionalization and Cokriging}\label{sec:lmc} 

Cokriging is the classical prediction method for multivariate spatial data. It uses a weighted average of existing data points to interpolate the values for the new location $s^*$. The weights are derived by minimizing the variance of estimation under an unbiased constraint. We will not show the full derivation here and refer for details for example to \cite{VerHoefCressie1993,Goovaerts1998,book_mult_geo}. Generally, locations further apart get less weight and the weights of the values at the $i$th location depend among other things on $C(s^*,s_i)$ which makes an estimate of the covariance structure needed for Cokriging. We want to emphasize here that the covariance between two locations is a matrix, hence a matrix-valued function needs to be modelled for Cokriging. This covariance function is usually estimated model-based by the popular linear model of coregionalization (LMC) \cite{lmc_paper,book_mult_geo}. The LMC model is defined via its covariance function
\[
	C(h) = \sum^r_{k=1}  T_k \rho_k(h)  , 
\]
where $T_k$ are positive semi-definite $p \times p$ matrices denoted as coregionalization matrices, $\rho_k$ are univariate, (parametric) spatial correlation functions and $r$ is the number of considered univariate correlation models. In practice, a family of univariate spatial correlation functions is selected, then the number $r$ is chosen and finally the unknown parameters of $\rho_k$ and the matrices $T_k$ are estimated.

\section{Neural Networks}\label{sec:nn}

Kriging and especially Cokriging are actually very involved prediction methods where the user has to choose data-specific covariance functions. These functions are usually chosen after visual inspections of the data and their estimated parameters. Neural networks on the other hand are universal function-approximators used in many different fields of data analysis with great success. In general, a fully connected neural network is a stack of layers where each layer consists of an affine transformation and an element-wise non-linear function that are applied on the input data to derive the output. The stack is created by using the output of one layer as an input to the following layer. A detailed discussion of neural networks is given for example in \cite{Goodfellow-et-al-2016}. 

In spatial data analysis, neural networks can be used for prediction tasks as a (non-parametric) alternative to Kriging, discarding the modelling of spatial cross-covariance \cite{neural_kriging, 4n_paper}. Naturally, the question arises which input features to use? A first option would be to use the coordinates of the field as input features and to use the values of the field as output. This idea has been used and extended for the univariate as well as the multivariate case in \cite{neural_kriging}.

\cite{4n_paper} discuss different ways of input feature selection for the univariate case. They propose a non-parametric version where the features consist of the coordinates on-site as well as the coordinates and values of the field for the $m$ nearest neighbours, with a total of $3m+2$ input features. Furthermore, they propose a parametric version where the input feature is solely the Ordinary Kriging prediction on-site and a unison of the parametric and non-parametric version with a total of $3m+3$ input features. This non-parametric feature selection approach from \cite{4n_paper} can be easily adapted to the multivariate case by simply considering the $p$ values of the random field for the $m$ nearest neighbouring points, which leads to a total of $m(p + 2) + 2$ input features and $p$ output features.

\section{Spatial Blind Source Separation}\label{sec:sbss}

Multivariate spatial models are usually defined via their covariance functions. Recently \cite{sbss_comp,10.1093/biomet/asz079} take a different approach and suggest a spatial blind source separation (SBSS) framework  by assuming that $X(s)$ can be formulated as a latent variable model of the form
$$
X(s) = \Omega Z(s) + \mu,
$$
where $\Omega \in \mathbb{R} ^ {p \times p}$ is a full-rank mixing matrix and $\mu$ is a $p$-variate location vector. The $p$-variate latent random field $Z(s)$ consists of $p$ independent, univariate random fields with $E(Z_i(s))=0$, $\var(Z_i(s))=1$ and $\cov(Z_i(s_1),Z_j(s_2)) = 0$ for all $i=1\dots,p$, $i \neq j$ and $s, s_1, s_2 \in \mathcal{S}$. 
\cite{sbss_comp,10.1093/biomet/asz079} show that the unmixing matrix and the latent fields can be recovered based on $X(s)$ alone (up to the usual BSS indeterminacies of signs and order) when each random field $Z_i(s)$ has a different covariance function $C_i(s_1,s_2) = \cov(Z_i(s_1),Z_i(s_2))$ for all $s_1, s_2 \in \mathcal{S}$. 
The strategy to find an unmixing matrix in \cite{sbss_comp,10.1093/biomet/asz079}
is to jointly diagonalize the regular sample covariance matrix and one or more so-called local covariance matrices. For a  centred  sample of size $n$, the local covariance matrices considered in \cite{10.1093/biomet/asz079} are of the form
 \[ 
 \hat{M}_f(X) = \frac{1}{n} \sum_{i=1}^n \sum_{j=1}^n f(s_i - s_j) X(s_i) X(s_j)^\top ,  
 \]
 where $f:\mathbb{R} ^ d \rightarrow \mathbb{R}$ is denoted as the kernel function that determines the locality of the covariance matrix. \cite{10.1093/biomet/asz079} consider three types of kernels, Gaussian, ball and ring, where the latter ones are of the form $f_b(h) = I(\| h \|  \leq r)$ and $f_r(h) = I( r_1 < \| h \|  \leq r_2 )$ respectively.
 
To estimate an unmixing matrix $\Gamma$, \cite{10.1093/biomet/asz079} suggest to firstly select $K$ non-overlapping sets of inner and outer ring radii $(r_{i1},r_{i2})$. Secondly, compute the $K$ corresponding local covariance matrices $\hat{M}_{f_i}(X^{st})$, $i=1,\ldots,K$ based on the whitened data $X^{st}=X^{st}(s)=\widehat \cov(X(s))^{-1/2}(X(s)-\bar X)$, where $\widehat \cov(X)$ denotes the sample covariance and $\bar X$ denotes the sample mean. Thirdly, find an orthogonal matrix $U$ which jointly diagonalizes all the $K$ local covariance matrices. In practice, this means that $U$ maximizes { \small $\sum_{i=1}^K \| \diag \left( U^\top \hat{M}_{f_i}(X^{st}) U \right) \|^2_F $ } where $ \| \cdot \|^2_F$ denotes the Frobenius norm. Many algorithms for joint diagonalization are available. We rely on one based on Givens rotations as described in \cite{10.2307/2347513}. The final estimate of the unmixing matrix $\Gamma$ is $\hat \Gamma = U^\top \widehat \cov(X(s))^{-1/2}$. For the properties and a profound discussion about SBSS see again \cite{10.1093/biomet/asz079}. Basically, SBSS and the estimator described here can be seen as an extension of SOBI \cite{Belouchrani:1997} for time series data to spatial data.
 
  
For multivariate spatial prediction the SBSS model is especially appealing as the cross-covariances of $Z(s)$ are zero. This leads to our proposed scheme of using SBSS as a pre-processing tool prior spatial prediction. In the first step SBSS as introduced in \cite{sbss_comp,10.1093/biomet/asz079} and summarized above is applied to estimate the spatially uncorrelated latent random field $Z(s)$ and the unmixing matrix $\hat \Gamma$. In the second step predictions for each entry of the latent field $Z(s)$ on some coordinate $s^*$ can be made independently by using Ordinary Kriging (or any other form of univariate spatial prediction). The final step leads to the prediction of the original field by mixing the predictions of the latent field with $\hat X(s^*) = \hat \Gamma^{-1} \hat Z(s^*) + \bar X$. We want to stress here that Ordinary Kriging (used here in Step 2) follows the same methodology as Cokriging described above, but the key difference is that only $p$ real-valued covariance functions have to be modelled, whereas in Cokriging the covariance function is $p \times p$ matrix-valued. This fact leads to a significant simplification in covariance modelling and hence multivariate spatial prediction in general.

\section{Simulation Study} \label{sec:simu}

As the proposed scheme of combining SBSS and Ordinary Kriging has not yet been explored we are interested to see if that procedure has an advantage over the more general Cokriging and if it can compete with predictions based on neural networks. For that purpose we conduct a simulation study comparing these methods in settings where the SBSS model does and does not hold. All the following simulations are carried out in R 3.5.1 \cite{r_language} using the packages SpatialBSS \cite{spatialbss_package}, RandomFields \cite{randomfields_package}, gstat \cite{gstat_package}, JADE \cite{jade_package} and keras \cite{keras_package}.

First we recall two popular univariate covariance functions. The Spherical covariance function has the from
\begin{small}
\[
C_{sp}(h; \sigma ^ 2, \phi) = \sigma ^ 2 \left( 1 - \frac{3}{2} \frac{h}{\phi} + \frac{1}{2} \left( \frac{h}{\phi} \right) ^ 3 \right) I\left( 0 \leq \frac{h}{\phi} \leq 1 \right),
\] 
\end{small}
where $\sigma ^ 2$ is a variance parameter and $\phi$ a range parameter. The M\'atern covariance function is given by
\begin{small}
\[
C_m(h; \sigma ^ 2, \nu, \phi) = \frac{\sigma ^ 2}{2 ^ {\nu - 1} \Gamma (\nu)} \left( \frac{h}{\phi} \right) ^ \nu  K_\nu \left( \frac{h}{\phi} \right),
\] 
\end{small}

where $K_\nu$ is the modified Bessel function of second kind with shape parameter $\nu$ and $\sigma ^ 2$ and $\phi$ are as above.

In the simulations, we consider three 3-variate random field settings. Setting 1 and Setting 2 follow a SBSS model with $C_1(h)=C_{sp}(h;1,2)$, $C_2(h)=C_{m}(h;1,0.5,2)$ and $C_3(h)=C_{m}(h;1,1,2)$ as covariance functions. In Setting 1 all three fields are Gaussian whereas in Setting 2 the three fields have a $t_5$ distribution. A $t_k$ distributed random field can be computed by simulating $k + 1$ centred Gaussian random fields $X_i(s)$ with unit variance and then apply $(X_1(s) \sqrt{k-1}) / ( (\sum_{i=2}^k X_i(s))^{1/2})$. The entries of the mixing matrix $\Omega$ are drawn from $N(0,1)$ for all simulations.
 
Setting 3 follows the parsimonious multivariate Mat\'ern model (PMat) as introduced in \cite{madoc31282}. In this model the marginals as well as the cross-covariances are Mat\'ern functions given by $C_{i}(h;\nu_i, \phi_i) =\sigma^2_{i} C_m(h; 1, \nu_i, \phi_i)$ for $i = 1, \dots, p$ and $C_{ij}(h;\nu_{ij}, \phi_{ij}) = \rho_{ij} \sigma_{i} \sigma_{j} C_m(h; 1, \nu_{ij}, \phi_{ij})$ for $i,j = 1, \dots, p, ~ i \neq j$ respectively. As derived in \cite{madoc31282} sufficient conditions for the parameters to provide a valid covariance model are $\phi_i = \phi_{ij} = \phi$, $\nu_{ij} = \frac{1}{2} (\nu_{i} + \nu_{j})$ for $i,j = 1, \dots, p$ and $\rho_{ij} = \beta_{ij} \frac{\Gamma (\nu_{i} + d/2) ^ {1/2} \Gamma (\nu_{j} + d/2) ^ {1/2} \Gamma (\nu_{ij}) }{\Gamma (\nu_{i}) ^ {1/2} \Gamma (\nu_{j}) ^ {1/2} \Gamma (\nu_{ij} + d/2)}$ for $i,j = 1, \dots, p, ~ i \neq j$, the matrix given by the values $\beta_{ij}$ being positive semi-definite with all diagonal elements equalling one and $d$ being the dimension of the domain of the random field. Hence, only the range parameter $\phi$, the elements $\beta_{ij}$ and the marginal shapes and variances are free parameters. We choose $\phi = 2$, $(\nu_{1}, \nu_{2}, \nu_{3}) = (0.25, 0.5, 1.0)$, $(\sigma^2_{1}, \sigma^2_{2}, \sigma^2_{3}) = (1, 1, 1)$ and $\beta_{ij} = 1$ for $i,j=1,\dots,3$.

For each of the three settings we consider two variants that differ in the way the sampling locations are selected. In the so-called uniform variant the x and y coordinates are independently sampled from a uniform $\mathcal{U} (0,1)$ distribution whereas in the skew variant the x and y coordinates are independently sampled from a beta $\mathcal{B} (2,5)$ and a uniform $\mathcal{U} (0,1)$ distribution respectively. Then all coordinates are multiplied by 35 to obtain the sampling domain $\mathcal{S}=[0,35] \times [0,35]$. Thus, for the uniform variant the points are uniformly scattered over the sampling domain whereas in the skew variant the focus is on the left part of the domain. We sample for each variant $n_{\mathcal{S}} = 1225$ sites and additionally define a grid $\mathcal{S}^* = \{ (x + 0.5, y + 0.5) : x, y \in \mathbb{Z} \cap [0,35] \}$ which again has $n_{\mathcal{S}^*} = 1225$ sites. For every iteration of the simulation we simulate the field for each coordinate variant on $\mathcal{S}$ and $\mathcal{S^*}$ at once and use for each coordinate variant the values of the field on $\mathcal{S}$ to predict the values on $\mathcal{S^*}$. Mean squared error (MSE) between the predictions and simulated (true) values on $\mathcal{S^*}$ are used as the performance measure of the spatial prediction, i.e. 
$\mbox{MSE} = 1/n_{\mathcal{S}^*} \sum_{\mathcal{S^*}} (X(s_i^*)-\hat X(s_i^*))^2$. In general, we expect the uniform variant to be easier than the skew variant. For each random field setting and coordinate variant we create 2000 replicate data sets. As for the selection of the tuning parameters and functions, we follow general guidelines from the literature in order to evaluate different methods.

\begin{table*}[t]
\small
\centering
\caption{Results of the different spatial interpolation models for all coordinate variants and random field settings. The mean(standard deviation) of the MSE for all 2000 simulation repetitions is presented.} 
\begin{tabular}{llccccc} \toprule
\textbf{Field Setting} 	& \textbf{Variant}   	& \textbf{BN} & \textbf{SNN}  & \textbf{LMC} &\textbf{SBSS-Kriging}  	& \textbf{SBSS-Kriging} 	 		\\ 
 & & & & \textbf{Cokriging} &\textbf{Spheric} & \textbf{Mat\'ern} \\ \midrule
SBSS Normal 			& uniform 				& 0.66(0.03)	& 0.63(0.03) & 0.57(0.03) 			& 0.49(0.02) 				& \textbf{0.47(0.02)}  			 \\ 
						& skew 					 	& 1.20(0.14)		& 1.16(0.13)		& 0.96(0.08) & 0.94(0.10) 				& \textbf{0.90(0.08)} 		 	 \\ 
 SBSS t5 				& uniform 				& 1.40(0.15)	& 1.32(0.15) 	& 1.26(0.15) 		& 1.10(0.13) 				& \textbf{1.07(0.13)} 			 \\
						& skew 				 	& 2.22(0.31)		& 2.13(0.29) 	& 1.82(0.21)	& 1.77(0.24) 				& \textbf{1.72(0.21)} 			 \\ 
PMat 					& uniform 				 	 & 0.34(0.02)		& 0.33(0.02)  & \textbf{0.29(0.01)}  & 0.30(0.02) 				& 0.31(0.02) 					 \\ 
						& skew 						& 0.56(0.07)		& 0.54(0.07) 	& 0.45(0.04) 	& \textbf{0.44(0.04)} 		& \textbf{0.44(0.04)}  					 \\ \bottomrule
\end{tabular}
\label{tab:results}
\end{table*}

The golden standard for multivariate spatial prediction is Cokriging with a LMC as covariance model. For the LMC we consider two univariate covariance functions with $r=2$, namely a Spherical covariance model and a nugget effect, which is an on-site variance term given by $C_n(h) = I(h=0)$, for details on the nugget effect see for example \cite{book_mult_geo}. For fitting the two coregionalization matrices $T_k$ we use an algorithm provided by the package gstat \cite{gstat_package}. This algorithm fits each covariance model individually and then finds the least squares closest to positive semi-definite matrix by setting negative eigenvalues to zero to find the matrices $T_k$, for details see the documentation of \cite{gstat_package}. As this algorithm needs the covariance function parameters prior fitting the coregionalization matrices we choose the range parameter $\phi$ of the Spherical covariance function to be the maximum distance between two points in the coordinate set divided by six. The motivation behind this value is that commonly the empirical semivariogram is evaluated up to a distance of a third of the maximum distance between two points in the set of coordinates and the vague assumption that spatial dependence reaches up to a third of the maximum considered semivariogram distance.

Additionally, we apply fully connected feed forward neural networks as a non-parametric model for spatial interpolation. We use the adaptation of the non-parametric approach of \cite{4n_paper} for the multivariate case as described above. Hence, the x and y coordinate values and the $p=3$ values of the field for the $m=10$ nearest neighbouring points as well as the on-site x and y coordinate values are considered, for a total of $52$ inputs features. The outputs of the network are simply the $p=3$ on-site field values. For the fully connected neural networks many options of activation functions and network structures are available. We rely on the popular Batch Normalization network (BN) \cite{Batch_norm} with Relu activation and additionally explore the use of Self-normalizing neural networks (SNN) \cite{NIPS2017_6698}. Hyper-parameter configurations are found by a grid search on one dataset for BNs and SNNs to find the five best working configurations, then each of the five best working configurations is freshly trained on every dataset of the simulations at hand and the best of the five configurations is used for prediction of the field on the grid. We use $80 \%$ of the available points $n_{\mathcal{S}}$ per dataset for training and the remaining for evaluation. The loss functions is the mean squared error (MSE) and we solely rely on the ADAM optimizer \cite{Adam_optim} with the proposed parameter setting. For the training we use an early stopping algorithm that stops training if the validation loss is not improving for $25$ consecutive training epochs. Closely following \cite{NIPS2017_6698} we use the following sets of hyper-parameters for the grid search: Number of hidden layers = $\{2,3,4,6,8 \}$, Number of nodes per layer = $\{8, 16, 32, 64, 128 \}$, L2 weight penalty = $\{ 0, 10^{-5}, 10^{-4}, 10^{-3} \} $, Mini batch size = $\{ 32, 64 \} $, Nodes progression = $\{ \mathrm{straight}, \mathrm{geometric}\} $ (straight denotes same number of nodes per layer and geometric denotes geometric progression of the first layers nodes to the output nodes) and additionally only for SNNs: Alpha dropout = $\{ 0.0, 0.05, 0.10 \}$.

As last prediction method, we use the formerly introduced three-step scheme of using SBSS as pre-processing in conjunction with Ordinary Kriging (SBSS-Kriging). We follow the recommendation of \cite{10.1093/biomet/asz079} and use numerous non-overlapping ring kernel functions. Specifically, we choose $K=4$ with the set of radii $\{(0,2), (2,4), (4,6), (6,8)\}$ to estimate the latent random field. To predict the values for each entry of the latent random field with Ordinary Kriging the covariance structures need to be estimated. Therefore, we choose two models. The first model is a Mat\'ern covariance function with an additional nugget effect for each entry of the latent field and the second model is a Spherical covariance function with an additional nugget effect also used for every entry of the random field. The parameters for these models are estimated with the algorithm implemented in the RandomFields package \cite{randomfields_package}. 

Table~\ref{tab:results} summarizes the MSE for the five considered spatial prediction methods for each of the three field settings for 2000 repetitions. The five methods in Table~\ref{tab:results} are listed by the order of the overall performance where the method performing best is highlighted. Overall LMC + Cokriging and SBSS-Kriging have very similar performance, when the SBSS model actually holds SBSS-Kriging slightly outperforms Cokriging. Also the Mat\'ern covariance matrix performs slightly better than the Spherical one. Spatial interpolation with neural networks shows inferior results compared to all Kriging methods and SNNs are slightly performing better than BNs. 

\begin{table*}[t]
\small
\centering
\caption{MSE for the various described spatial prediction methods for the moss dataset. 20\% (119) and 80\% (475) of the data are used for testing and training.} 
\begin{tabular}{ccccccccc} 
\toprule
\textbf{BN5} & \textbf{BN10} & \textbf{SNN5} & \textbf{SNN10} & \textbf{LMC} & \textbf{SBSS-Kriging}	& \textbf{SBSS-Kriging} & \textbf{SBSS-Kriging} 	& \textbf{SBSS-Kriging} \\ 
& & & & \textbf{Cokriging} & \textbf{Ball Spheric}	& \textbf{Ball Mat\'ern} & \textbf{Ring Spheric}	& \textbf{Ring Mat\'ern} \\ \midrule
0.173 & 0.173 & 0.207 & 0.191 & 0.134 & 0.129 & 0.129 & \textbf{0.128} & \textbf{0.128} \\ \bottomrule
\end{tabular}
\label{tab:results_kola}
\end{table*}

\section{Real Data Example}\label{sec:real_data}
In order to test the prediction power of the above discussed methods we consider a geochemical dataset. Specifically, we look at the concentration of 31 chemicals (Ag, Al, As, B, Ba, Bi, Ca, Cd, Co, Cr, Cu, Fe, Hg, K, Mg, Mn, Mo, Na, Ni, P, Pb, Rb, S, Sb, Si, Sr, Th, Tl, U, V and Zn) measured in samples of terrestrial moss collected at 594 locations in Norway, Finland and Russia alongside the Barent sea. This moss dataset collected in the Kola project is freely available in the R package StatDA \cite{statda_package} and is described in detail in \cite{moss_book}. Moreover, it was already considered in the context of SBSS in \cite{sbss_comp,10.1093/biomet/asz079}. Closely following the former publications this geochemical dataset is of compositional nature, meaning that the information lies in the relative values rather than in the absolute ones. We respect this fact by transforming the original values into ilr (isometric log-ratio) coordinates by using pivot ilr coordinates and carry out all of the following analysis in that space. Details of compositional data as well as the according transformations can be found in for example \cite{enlighten33520}, details of compositional data in the context of SBSS are described in \cite{sbss_comp}. Note that the ilr transformation reduces the dimension of the data in this case from $31$ to $30$, hence $p=30$. To test the prediction power of the various spatial prediction methods we randomly put aside 20\% (119) of the sample locations and use the remaining 80\% (475) as a training set. The training set is used to train the various spatial interpolation methods. Trained models are used to predict the $30$ field values on the remaining sites. As a final step the MSE is computed between the predictions and the values of the test set. 

For neural networks we again consider Batch Normalizing as well as Self-normalizing neural networks in conjunction with the nearest neighbour input feature selection described above, hence, a total of $m(p + 2) + 2$ inputs and $p$ outputs are used. As the dataset contains less samples as were considered in the simulation study we take $m=10$ neighbours (BN10, SNN10) and additionally we also explore the use of $m=5$ neighbours (BN5, SNN5) leading to $322$ and $162$ input features respectively. Training and searching for hyper-parameters is done in the same fashion and with the same hyper-parameter as in the simulation study.
We also carry out Cokriging with the use of the LMC with a nugget and a Spherical covariance model in the same fashion as in the simulation study.
For the SBSS-Kriging approach we use again the Mat\'ern and Spherical model with a nugget effect to derive the Ordinary Kriging predictions on each entry of the latent field individually. The actual estimation of the latent field with SBSS needs an educated choice of the considered kernel functions. \cite{sbss_comp} came to the conclusion that a ball kernel with a radius of 50 km is the appropriate choice. \cite{10.1093/biomet/asz079} found that the result of SBSS is more stable when using numerous non-overlapping ring kernel functions in general and used various such settings for the moss dataset. Based on these considerations we carry out SBSS with $K=1$ ball kernel with a radius of 50 km and also use $K=4$ non-overlapping ring kernels with the following set of radii in km $\{(0,25), (25,50),$ $ (50,75), (75,100)\}$. This leads to a total of four different SBSS-Kriging models.

The MSE for the nine formerly discussed spatial prediction methods for the moss dataset are summarized in Table~\ref{tab:results_kola}. Similar as in the simulation study the results show that SBSS-Kriging seems to deliver comparable results to the other considered spatial prediction methods and neural networks seem to perform inferior. Moreover, the two spatial kernel functions settings for SBSS seem to perform equally.

\section{Discussion and Conclusion}

In this paper we give an overview of different methods for multivariate spatial prediction, the golden standard method Cokriging with an LMC covariance function in Section~\ref{sec:lmc}, neural networks in Section~\ref{sec:nn} and we present how SBSS (introduced in \cite{sbss_comp,10.1093/biomet/asz079}) can be used to reduce the task of multivariate spatial prediction to several univariate ones in Section~\ref{sec:sbss}. To the best of our knowledge SBSS has not yet been used in this context. To investigate the effectiveness of this procedure we carry out an extensive simulation study where different random field models are simulated and predicted in Section~\ref{sec:simu}. Table~\ref{tab:results} shows the simulation results and hints that SBSS-Kriging performs at least equally as Cokriging, and as expected performs best when the SBSS model actually holds. As SBSS-Kriging performs comparable with Cokriging it seems to be generally favourable when considering the fact that only univariate covariance functions have to be modelled. Although neural networks show state-of-the-art performance in many different tasks, our simulations indicate that neural networks are a valid tool for spatial interpolation but are not able to produce comparable results as the various Kriging methods considered here. On the other hand neural networks could also be used in favour of Ordinary Kriging when SBSS is used prior spatial prediction, which is a starting point for further research. Naturally, many other settings could have been chosen for comparison but we believe this gives a first idea about the performance of these multivariate prediction methods. Moreover, we confirm the findings from the simulation study by applying all the considered spatial interpolation methods to a geochemical dataset in Section~\ref{sec:real_data}.

\section{Acknowledgements}

This work was supported by the Austrian Science Fund P31881-N32.

\bibliographystyle{unsrt}


\end{document}